\documentclass{ws-procs961x669}            
\begin{document}

\newcommand{\rma}{\rho_m}
\newcommand{\rr}{\rho_r}
\newcommand{\rL}{\rho_\Lambda}
\newcommand{\PL}{P_\Lambda}
\newcommand{\dG}{\dot{G}}
\newcommand{\drr}{\dot{\rho}_r}
\newcommand{\drL}{\dot{\rho}_\Lambda}
\newcommand{\dH}{\dot{H}}
\newcommand{\CC}{\Lambda}

\newcommand{\rv}{\rho_{\rm vac}}
\newcommand{\Pv}{P_{\rm vac}}
\newcommand{\rvo}{\rho^0_{\rm vac}}

\newcommand{\OM}{\Omega_M}
\newcommand{\Om}{\Omega_m}
\newcommand{\Omo}{\Omega^0_{m}}
\newcommand{\Omh}{\hat{\Omega}_m}
\newcommand{\OMo}{\Omega_{M}^0}
\newcommand{\ORo}{\Omega_{r}}
\newcommand{\OL}{\Omega_{\rm vac}}
\newcommand{\Oc}{\Omega_{c}}
\newcommand{\Oco}{\Omega_{c 0}}
\newcommand{\OLh}{\hat{\Omega}_{\Lambda}}
\newcommand{\GL}{\Gamma_{\Lambda}}
\newcommand{\GLh}{\hat{\Gamma}_{\Lambda}}
\newcommand{\OLo}{\Omega^0_{\rm vac}}
\newcommand{\OX}{\Omega_{X}}
\newcommand{\OXo}{\Omega_{X}^0}
\newcommand{\OXh}{\hat{\Omega}_{X}}
\newcommand{\OD}{\Omega_{\rm DE}}
\newcommand{\ODo}{\Omega_{\rm DE}^0}
\newcommand{\OR}{\Omega_R}
\newcommand{\OK}{\Omega_K}
\newcommand{\OKo}{\Omega_{K}^0}
\newcommand{\OZ}{\Omega_0}
\newcommand{\OT}{\Omega_T}
\newcommand{\rc}{\rho_c}
\newcommand{\rco}{\rho^0_{c}}
\newcommand{\rmo}{\rho_{m 0}}
\newcommand{\rs}{\rho_s}
\newcommand{\ps}{p_s}
\newcommand{\rM}{\rho_m}
\newcommand{\rmr}{\rho_m}
\newcommand{\pmr}{p_m}
\newcommand{\rMo}{\rho_{m}^0}
\newcommand{\pM}{p_m}
\newcommand{\rR}{\rho_r}
\newcommand{\rD}{\rho_{\rm DE}}
\newcommand{\rDt}{\tilde{\rho}_D}
\newcommand{\rDo}{\rho_{D}^0}
\newcommand{\rX}{\rho_X}
\newcommand{\pX}{p_X}
\newcommand{\wX}{w_X}
\newcommand{\wm}{\omega_m}
\newcommand{\wR}{\omega_R}
\newcommand{\aR}{\alpha_R}
\newcommand{\amr}{\alpha_m}
\newcommand{\aef}{\alpha_e}
\newcommand{\aX}{\alpha_X}
\newcommand{\rLo}{\rho^0_{\CC}}
\newcommand{\pD}{p_D}
\newcommand{\wD}{\omega_D}
\newcommand{\zm}{z_{\rm max}}
\newcommand{\wL}{\omega_{\CC}}
\newcommand{\CCo}{\Lambda_0}
\newcommand{\we}{\omega_{e}}
\newcommand{\re}{r_{\epsilon}}
\newcommand{\tOM}{\tilde{\Omega}_M}
\newcommand{\tOm}{\tilde{\Omega}_m}
\newcommand{\tOmo}{\tilde{\Omega}_m^0}
\newcommand{\tOL}{\tilde{\Omega}_{\CC}}
\newcommand{\tOD}{\tilde{\Omega}_{D}}
\newcommand{\tODo}{\tilde{\Omega}_{D}^0}
\newcommand{\xL}{\xi_{\CC}}
\newcommand{\fM}{f_{M}}
\newcommand{\fL}{f_{\Lambda}}
\newcommand{\lu}{\lambda_1}
\newcommand{\ld}{\lambda_2}
\newcommand{\lt}{\lambda_3}
\newcommand{\model}{X$\CC$CDM}
\newcommand{\f}{\tilde{f}}
\newcommand{\cM}{{\cal M}}
\newcommand{\cMd}{{\cal M}^2}
\newcommand{\ka}{\kappa}

\newcommand{\bCC}{\bar{\CC}}
\newcommand{\bCDM}{\bar{\CC}{\rm CDM}}

\newcommand{\CH}{C_H}
\newcommand{\CHd}{C_{\dot{H}}}
\newcommand{\rDE}{\rho_{\rm DE}}
\newcommand{\tetm}{\theta_{\rm m}}
\newcommand{\rplu}{r_{+}}
\newcommand{\rmin}{r_{-}}
\newcommand{\nueff}{\nu}
\newcommand{\nueffp}{\nu_{\rm eff}'}
\newcommand{\xim}{\xi_m}
\newcommand{\xiR}{\xi'}
\newcommand{\rRo}{\rho_{r 0}}
\newcommand{\bk}{{\bf k}}
\newcommand{\mpl}{m_{\rm Pl}}
\newcommand{\MPl}{{\cal M}_{\rm Pl}}

\newcommand{\be}{\begin{equation}}
\newcommand{\ee}{\end{equation}}

\newcommand{\pL}{p_{\CC}}

\newcommand{\wXCDM}{$w${\rm XCDM}\,}
\newcommand{\CCS}{\CC_s{\rm CDM}}
\newcommand{\wY}{w_Y}


\newcommand{\cH}{\mathcal{H}}
\newcommand{\cpH}{\mathcal{H}^\prime}
\newcommand{\cpHs}{\mathcal{H}^{\prime 2}}
\newcommand{\cppH}{\mathcal{H}^{\prime \prime}}
\newcommand{\txi}{\tilde{\xi}}
\newcommand{\ha}{\hat{a}}
\newcommand{\astar}{a_{*}}
\newcommand{\trI}{\tilde{\rho}_I}
\newcommand{\rI}{\rho_I}
\newcommand{\TI}{T_I}
\newcommand{\tHI}{\tilde{H}_I}

\hyphenation{theo-re-ti-cal gra-vi-ta-tio-nal theo-re-ti-cally tu-ning}


\title{Composite  running vacuum in the Universe: implications on the cosmological tensions}

\author{Joan Sol\`a Peracaula$^{a,b}$\footnote{Invited talk  at MG 17, Pescara, Italy, July 7-12, 2024}}

\address{$^a$Departament de F\'isica Qu\`antica i Astrof\'isica, \\
and \\ $^b$ Institute of Cosmos Sciences,\\ Universitat de Barcelona, \\
Av. Diagonal 647, E-08028 Barcelona, Catalonia, Spain,
\vspace{0.1cm}\\
$^*$E-mail: sola@fqa.ub.edu}

\begin{abstract}
The possibility that the vacuum energy density (VED) could be time dependent in the expanding Universe is intuitively more reasonable than just a rigid cosmological constant for the entire cosmic history. The framework of the running vacuum model (RVM) is a salient example derived from QFT in curved spacetime, wherein the VED appears as a power series of the Hubble rate, $H(t)$, and its derivatives. The RVM contributes to alleviate the cosmological tensions and at a more fundamental level it also helps to smooth out certain hardcore aspects of the cosmological constant problem. Composite dark energy (DE) extensions of the RVM are possible, in which the DE  is a mixed fluid made out of running vacuum and an entity X called ``phantom matter'' which, I should stress, is radically different from phantom DE, since the former produces positive pressure like ordinary matter (therein its name). The prototype is the old $\Lambda$XCDM model\cite{Grande:2006nn}. Recently a simplified version of the latter, called  the $w$XCDM\cite{wXCDM},  has proven capable to yield a truly outstanding fit to the cosmological data with  dramatic implications for mitigating the cosmological tensions. The analysis is sensitive to the kind of BAO data used, transversal (2D) or anisotropic (3D).  For both BAO types the overall fit quality  is substantially better than for the standard $\Lambda$CDM model, and the growth tension becomes alleviated. Nevertheless, only for  2D BAO the $H_0$-tension can be completely cut down. Most notably, this scenario favours  (at $3.3\sigma$ c.l.) quintessence-like behavior around our time, as recently reported by the DESI Collaboration. In what follows I will summarize in very simple phenomenological terms  the essentials  of the RVM idea and its composite extensions,  as well as the  remarkable practical implications ensuing from such scenarios.
\end{abstract}


\bodymatter

\section{Introduction}\label{sec:intro}

Despite the standard or concordance model of cosmology, aka $\CC$CDM\,\cite{Peebles1984,Peebles1993,KraussTurner1995,OstrikerSteinhardt1995},  has proven to be a rather successful  theoretical framework for the description of the Universe in the last few decades, unfortunately the model  entails a number of serious pitfalls which raise serious doubts about its structure and viability. Dark matter (DM), for example,  has never been found thus far, a fact which is extremely worrisome as otherwise we cannot understand the dynamical origin of the  large scale structure (LSS) that we observe\,\cite{Peebles1993}.   On the theoretical side, the core ingredient of the model, namely the cosmological constant (CC), $\CC$,  is an unending source of headaches for the theoretical physicists. While the $\CC$-term in the gravitational  field equations was introduced by A. Einstein 107 years ago\,\cite{Einstein1917},  the potential existence of a big conundrum associated with it, the so-called  `cosmological constant problem' (CCP), was unveiled 50 years later by Y. B. Zeldovich\,\cite{Zeldovich1967}.  The CCP is the chocking  realization that the manyfold successes of QFT in the world of the elementary particles turn into a blatant fiasco in the realm of gravity. This is because  QFT predicts a value for $\rv$ which is excruciatingly much larger than that of the current critical density of the universe and that of matter at present $\rho_m^0$\,\cite{Weinberg89,PeeblesRatra2003,Padmanabhan2003,Copeland2006,JSPRev2013,JSPRev2015,JSPRev2022}\,\footnote{For an informal, but rather vivid, introduction to the Cosmological Constant Problem, see e.g. \cite{JSPCosmoverse}.}.   Possible dynamical solutions to the CCP were approached phenomenologically  within classical field theory  trying to explain  the value of  $\rv$ in terms of the effective potential of a cosmic scalar field which settles down dynamically (hence without fine tuning) to the present current value, $\rvo\sim 10^{-47}$ GeV$^4$, see e.g.\,\cite{Dolgov82,Abbott85,Banks85,PSW87,Barr87,Ford87,Sola89}. A well-known example is the ``cosmon model'' put forward in \cite{PSW87}, which was further discussed by Weinberg  in his famous review\,\cite{Weinberg89}.   Later on, the notion of quintessence broke through and the first models appeared in two different waves separated by a full decade\cite{Wetterich88,RatraPeebles88,Caldwell98}. Here the purpose was much more modest. Rather than trying to explain the value of $\rvo$ (the hardcore aspect of the CCP, i.e. the  ``old cosmological constant problem''\cite{Weinberg89}), they aimed at finding an explanation for a lesser problem,  the so-called  ``cosmic coincidence problem''\cite{Steinhardt1997}, although it is not even clear if such a  `cosmic coincidence' ($\rho_m^0\simeq \rvo$)  is a problem at all\,\cite{Ishak2005}.

The $\CC$CDM cannot account for these fundamental problems, not even for more mundane ones. It  is formulated in  the context of the Friedman-Lemaitre-Robertson-Walker (FLRW) framework and is deeply ingrained in the General Relativity (GR) paradigm.  However, one of the most important drawbacks of GR  is that it is a non-renormalizable theory.  This can be considered a serious theoretical obstruction for GR to be considered a fundamental theory of gravity. Obviously, this fact  impacts negatively on the $\CC$CDM too.  As a result GR cannot properly describe the short distance effects of gravity, i.e.  the ultraviolet regime (UV) prevailing in the very early universe, only the large distance effects (or infrared regime) characteristic of the current universe. This means that GR cannot provide by itself a framework for quantizing gravity (the spacetime metric field)  along with the rest of the fundamental  interactions.  Notwithstanding, if one treats gravity as a classical (external or background) field we can still quantize the matter fields of the fundamental interactions (electroweak and strong interactions). This is the main program of the semiclassical approach, namely  the point of view of QFT in curved spacetime\,\cite{BirrellDavies82,ParkerToms09,Fulling89}.   Despite the time-honored status of such a semiclassical approach,  a proper renormalization of the VED  within  QFT in curved spacetime with practical implications on the problems of modern cosmology was not accomplished until more recently through the idea of running vacuum\,\cite{JSPRev2022}, see also \cite{JSPRev2013,JSPRev2015}.   The running vacuum model (RVM) indeed leads to an effective form of the $\CC$CDM (which was called $\overline{\Lambda}$CDM in \cite{JSPRev2015}),  in which the physical value of $\CC$ (which is not just the parameter in the action) `runs' smoothly with the cosmic expansion  (represented  by the Hubble rate $H$)  thanks to the quantum matter effects.  This running, in fact, can be thought of as a renormalization group running\,\cite{JSPRev2013}. As a result,  the quantum vacuum  does not remain static throughout the cosmic  evolution. To put it crystal-clear: there is no such thing as a `cosmological constant' in the QFT context.  The running nature of the vacuum has also been substantiated in the framework of low-energy effective
string theory\,\cite{ReviewNickJoan2021,PhantomVacuum2021}, see also \cite{BasMavSol,NickPhiloTrans,Gomez-Valent:2023hov,Dorlis2024} and the forthcoming comprehensive review \cite{NickJoan_PR}. Despite the deviation of the physical $\CC$ from constancy  is, of course,  small since it originates from quantum effects, the consequences on the  $\CC$CDM behavior with a running $\CC$ need not be negligible; in fact, they can be essential to cure the tensions.  In point of fact, in the RVM framework, the mild dynamics of  the VED -- and hence the running of the physical $\CC$ --  emerges from first principles (viz. QFT and string theory). One finds  that the evolution appears through powers of the expansion rate $H(t)$ and its derivatives, the leading correction being of order $H^2(t)$ \footnote{Together with the approach to the RVM from QFT in curved spacetime and string theory, recently a completely different strategy to the study of the cosmological vacuçum energy based on  lattice Quantum Gravity (QG)\,\cite{Dai:2024vjc} was able to  derive  the same kind of  dynamical effects  $\sim H^2$  on the VED that are characteristic of the RVM for the late universe\,\cite{JSPRev2022}. The three approaches, therefore, concur in a same dynamical VED structure, which is remarkable.}.  In addition, and in fact most notably, the vacuum evolution in the  RVM context  proves  to be completely free from the quartic mass effects  $\sim m^4$  which are brandished  too often   in the context of simplified (in fact, wrong)  renormalization treatments of the VED.  Thus, the RVM may provide a possible solution to the excruciating fine tuning problem inherent to the CCP\,\cite{CristianJoan2020,CristianJoan2022a,CristianJoan2022b,CristianJoanSamira2023}.  At the end of the day,  the RVM formulation of QFT in curved spacetime  raises a very serious question for our consideration, to wit:   is the CCP still in force in the running vacuum theoretical framework?   Once more I refer the reader to the simplified exposition \cite{JSPCosmoverse}, if she/he does not wish to go through the previous technical papers.

The RVM formulation does not only have a bearing on high brow fundamental theoretical problems such as the CCP, it also impinges on very practical --  in fact,  ground level --  matters of modern cosmology, such as  a possible resolution of the current cosmological tensions between the concordance $\CC$CDM model and different sorts of data. Recall that the tensions are concerned, in particular,  with the measurement of the current Hubble parameter $H_0\equiv 100 h$ km/s/Mpc  ($h\simeq 0.7$) and the growth of  LSS structures.  The growth rate is usually gauged by means of the parameters $S_8$ or $\sigma_8$, or even better with $\sigma_{12}$, which defines  the amplitude of the matter power spectrum at fixed spheres of radius $12$ Mpc rather than $8h^{-1}$ Mpc. In this way one can avoid artificial dependence on the value of $h$\,\cite{Sanchez:2020vvb,eBOSS:2021poy,Semenaite:2022unt}. The $H_0$-tension involves a  serious disagreement  between the value of $H_0$ inferred from  CMB observations, making use of fiducial $\CC$CDM cosmology, and the corresponding value extracted from the local (distance ladder) measurements.  Comparison between the two types of measurements  leads to a severe inconsistency of $\sim 5 \sigma$ c.l. The growth tension, on the other hand,  is related with the overproduction  of large scale structure in the late universe as predicted by the $\CC$CDM in comparison to actual measurements. The discrepancy here is moderate but persistent,  at a level of  $\sim 2-3\sigma$.  See  \cite{Perivolaropoulos:2021jda,Abdalla:2022yfr} for comprehensive reviews on these tensions, and \cite{Vagnozzi:2019ezj}  and references therein  for the impact of new physics.    Furthermore, a new kind of (severe) tension has popped up more recently in the landscape of the $\CC$CDM. It stems  from the observations by the  James Webb Space Telescope (JWST)\,\cite{Gardner:2006ky,Labbe:2022ahb}, which  have revealed the existence of an unexpectedly large population of supermassive galaxies at large redshifts in the approximate range  $z\gtrsim 5-10$, which is completely at odds within the  expectations of the concordance $\CC$CDM model.

A wide panoply of  strategies in the literature have been proposed  to alleviate some of the above tensions. For instance, it has been argued that within the class of models where the DE is dealt with as a cosmic fluid with equation of state (EoS) $w(z)$, solving the $H_0$ tension demands the phantom condition $w(z)<-1$ at some $z$, whereas solving both the $H_0$ and $\sigma_8$ tensions requires $w(z)$ to cross the phantom divide and/or other sorts of exotic transitions, see e.g. \cite{Heisenberg:2022gqk,VariousTransitions,Gomez-Valent:2023uof}. Most conspicuously, the possibility of a sign flip of the $\CC$ term has been entertained in recent times.  Particularly the  $\CCS$ model,  analyzed in \cite{Akarsu:2021fol,Akarsu:2023mfb} (based on the framework of \cite{Akarsu:2019hmw}), in which one considers a sudden transition from anti-de Sitter  (AdS), hence $\CC<0$, into de Sitter (dS)  regime ($\CC>0$) occurring near our time. Transversal BAO is employed in the fitting analysis, see also \cite{Gomez-Valent:2023uof}. Nonetheless, a more general model ($w$XCDM) has been proposed in \cite{wXCDM} which makes the fit more robust and  manifestly more consistent with recent observations, for  it leads to quintessence behavior near our time (rather than  $\CC=$const. as in the $\CCS$). Thus, $w$XCDM fits in with the recent DESI data pointing to quintessence\cite{DESI:2024mwx}.  Model  $w$XCDM is inspired in the old $\CC$XCDM model\,\cite{Grande:2006nn,Grande:2006qi,Grande:2008re}, a composite DE model in which  the running vacuum and an extra $X$ component (playing the role of `phantom matter')  can exchange energy.

Obviously, more studies will be needed before getting a final confirmation of the dynamical nature of the DE and its composite structure. It  is, however, remarkable to note the existence  of a fairly amount of  harbinger work in the literature pointing to the dynamics of the DE about a decade ago, which was based on the analysis of a significant amount of cosmological data. The evidence collected ranged in between $3-4\sigma$, see\,\cite{Gomez-Valent:2014rxa,Sola:2015wwa,Sola:2016jky,Sola:2017znb,SolaPeracaula:2016qlq,SolaPeracaula:2017esw}. These analyses  involved the running vacuum model (RVM).  See also the subsequent works using different approaches\,\cite{Zhao:2017cud,SolaPeracaula:2018wwm}.

 Finally, we point out that  the dynamical nature of the quantum vacuum can provide a fruitful arena for the potential time variation of the fundamental constants\,\cite{Fritzsch:2012qc}, which is  a very active field of research that could provide new evidences on the nature of the DE, see e.g. \cite{MemorialHF2024} and references therein.  In what follows, I will mainly concentrate on how the running of  the VED can  help to cure the cosmological  tensions and on the important role that the  composite extensions of the RVM may play in the description of the cosmological data.

\section{Canonical RVM and composite RVM}\label{sec:LXCDM}
In what follows I will summarize the idea behind the original running vacuum model (RVM) and its composite extension, the $\CC$XCDM model.

\subsection{RVM}
As promised, I shall skip all technical QFT details here --  see \,\cite{CristianJoan2020,CristianJoan2022a,CristianJoan2022b,CristianJoanSamira2023} or the reviews \cite{JSPRev2022}  in case you are interested.  Here only  phenomenological aspects will be touched upon.   I will say though only one qualitative technicality, which is important enough:  the VED that we measure today through the physical cosmological term, namely  $\rvo=\Lambda_{\rm obs}/(8\pi G_N)$, is not just the bare term $\rL=\Lambda_b/(8\pi G_b)$ that we may construct from the bare CC value $\Lambda_b$ (and the bare Newton's coupling $G_b$)  in the action, not even the one that we may construct from the renormalized value $\rL (M) $   (which depends on some appropriate renormalization scale $M$).  To construct the physically renormalized VED we have to add the renormalized $\rL (M) $  and the renormalized zero-point energy (ZPE) in the corresponding renormalization scheme.  We may write this very qualitatively  as follows:  ${\rm VED}=\rho_\Lambda+{\rm ZPE}$.  Hence the physical cosmological term that we measure is $\Lambda_{\rm obs}=(8\pi G_N) {\rm VED}\equiv (8\pi G_N) \rvo$.   It is thanks to the presence of the ZPE in curved spacetime that the VED acquires a dependence on the expansion rate $H$ and therefore $\Lambda_{\rm obs}=\Lambda_{\rm obs}(H)$  becomes   a dynamical quantity in an expanding universe. I repeat: there is no such thing as a rigid cosmological constant in the quantum field theory  of an expanding universe!   Previous approaches to the subject over the years  failed blatantly in realizing this crucial fact.

Indeed,  the  renormalized result still requires a  physical interpretation since it depends on the renormalization scale $M$. The vacuum effective action $W_{\rm eff}$\,\cite{BirrellDavies82} is explicitly dependent on the renormalization scale $M$ despite the fact that the  full effective action is not.  An adequate choice of $M$ at the end of the renormalization program  obtains if one picks  it equal to the value of $H$ at each cosmic epoch under consideration. This  corresponds to choose the renormalization group scale around the characteristic energy scale of  FLRW spacetime at any given moment, and hence it should have physical significance.  This is similar to the standard practice in ordinary gauge theories, where the choice of  the renormalization scale  is  made near the typical energy of the process.  As previously indicated, the setting  $M=H$ has also been recently confirmed as being  fully consistent within the   lattice QG approach of\,\cite{Dai:2024vjc}.  It follows that the renormalized VED that we find is an evolving quantity with the cosmic expansion (see the above mentioned papers).  The simplest situation which allows to illustrate this fact  is by considering a quantized scalar field $\phi$ with mass $m_\phi$ endowed  with a non-minimal coupling to gravity (i.e. having a term $\sim\xi\phi^2R$ in the action, $\xi$ being the non-minimal coupling)\footnote{One can generalize the QFT computations for an arbitrary number of quantized scalar fields and even an arbitrary number of quantized fermions fields, see\,\cite{CristianJoanSamira2023}, but for the sake of simplicity  it will suffice to describe here  the case with one single scalar field. The form \eqref{VacuumEnergyDensityScalarField} of the VED is the same in all cases. The differences are encoded in the structure of the effective parameter $\nueff$. }.  Its quantum matter fluctuations in the expanding FLRW background produce an evolution of the VED with the cosmic expansion.
The final result for the running  VED at low energies  between two expansion history times, say  the current epoch (characterized by the value $H_0$ of the Hubble parameter) and  some nearby epoch $H$ of the cosmic evolution can be rendered in the following  very compact form (see the aforementioned papers):
\begin{equation}\label{VacuumEnergyDensityScalarField}
\rho_{\rm vac}(H)=\rho_{\rm vac}^0+\frac{3\nueff}{8\pi}\left(H^2-H_0^2\right)\mpl^2+{\cal O}(H^4),
\end{equation}
where the leading form of $\nueff$ is constant (with only a mild log evolution with $H$\cite{CristianJoan2022a}) and reads
\begin{equation}\label{eq:nueffBososns}
\nueff\equiv \frac{1}{2\pi}\left(\xi-\frac{1}{6}\right)\frac{m_\phi^2}{\mpl^2}\ln \frac{m_\phi^2}{H_0^2}\,,
\end{equation}
 $\rho_{\rm vac}^0 \equiv \rho_{\rm vac} (H_0)$ being the current value of the VED (accessible from  observations) and  $H_0$   today's value of the Hubble function. It is necessary to remark that $\nueff$ is an effective parameter expected to be small due to its proportionality to $m_\phi^2 / \mpl^2$.  The higher powers of $H$, indicated in the above equations as ${\cal O}(H^4)$,  are not considered here since they can be significant only for the physics of the  very early universe, namely during the inflationary time, and hence the above low-energy formula applies virtually to any (post-inflationary) epoch. The above dynamical expression for the VED turns out to adopt the canonical RVM form, see\,\cite{,JSPRev2013,JSPRev2015,JSPRev2022} and references therein.  This formula proves rather successful from the phenomenological point of view.  Fitting the overall cosmological data with it  indeed provides an estimate for $\nueff$ at the level of  $\nueff \sim 10^{-4}-10^{-3}$\,\cite{SolaPeracaula:2021gxi,SolaPeracaula:2023swx}, see also \cite{Gomez-Valent:2014rxa,Sola:2015wwa,Sola:2016jky,SolaPeracaula:2016qlq,Sola:2017znb,SolaPeracaula:2017esw}. Such a  phenomenological determination of $\nueff$   lies in the ballpark of old theoretical expectations\,\cite{Sola:2007sv} and is compatible with the recent lattice QG approach of\,\cite{Dai:2024vjc}. Furthermore, the order of magnitude of $\nueff$  is reasonable if we take into account that the masses involved here pertain of course to the scale of a typical Grand Unified Theory (GUT) where, in addition, a large factor must be included to encompass the  large multiplicity of heavy particles.  It is also worth noticing that the obtained order of magnitude for $\nueff$  is  compatible with the Big Bang nucleosynthesis (BBN)  bound\cite{Asimakis:2021yct}.

An important quantity is the effective equation of state (EoS) for the running vacuum.  Its study requires the computation of the quantum effects on the vacuum pressure, which produce a deviation from the classical result $\Pv=-\rv$.  This is remarkable as it enables the quantum vacuum to mimic quintessence and phantom DE.  The leading expression of the EoS  for the current universe is the following\cite{CristianJoan2022a,CristianJoan2022b}:
\begin{equation}
w_{\rm vac}=\frac{P_{\rm vac}(H)}{\rho_{\rm vac}(H)} \approx -1-\nueff\frac{\dot{H}\mpl^2}{4\pi \rho^0_{\rm vac}}\,,
\end{equation}
where again the ${\cal  O}(H^4)$ effects are omitted. For very low redshift  $z$ and in terms of the current cosmological parameters $\Omega^0_i=\rho^0_i/\rho^0_c=8\pi G_N\rho^0_i/(3H_0^2)$ the above expression reduces to\cite{CristianJoan2022b}
\begin{equation}\label{EqStateScalar}
w_{\rm vac}(z)\approx -1+\nueff\frac{\Omega_{\rm vac}^0}{\Omega_{\rm m}^0}(1+z)^3.
\end{equation}
This result is worth emphasizing since it quantifies the small departure  of the vacuum EoS from -1. Since  $\nueff \sim 10^{-4}-10^{-3}$ is not that small this EoS correction could perhaps be measured in the late universe. For positive (resp. negative)  sign of $\nueff$,   Eq.\,\eqref{EqStateScalar} predicts that the vacuum energy behaves as quintessence (resp. phantom DE) for small $z$. From this point of view, a departure of the effective EoS of the DE from $-1$ does not necessarily imply the existence of quintessence or phantom fields, as the (quantum) vacuum itself can be responsible for these effects!  This is a remarkable qualitatively new feature of the quantum vacuum in the RVM formulation.

As noted before, the EoS expression\eqref{EqStateScalar} is valid only for small values of the redshift $z$. One  can show that the deviation is even bigger in the remote past  (although the exact formula is more complicated), adopting a kind of `chameleonic' behavior by which the EoS of the quantum vacuum tracks the EoS of matter at high redshifts, see\,\cite{CristianJoan2022b,Moreno-Pulido:2023drk} for more details.  It is important to stress that these  results have been obtained from first principles, namely from explicit QFT calculations in the FLRW background, hence they rely on the properties of the quantum vacuum. There is therefore no need to introduce \textit{ad hoc} fields to make the DE dynamical: the quantum vacuum is already so!

\subsection{Composite running vacuum: the $\CC$XCDM}
Next we move to a composite RVM scenario, the $\CC$XCDM.   It was proposed for the first time in \cite{Grande:2006nn}  and further expanded in subsequent works \cite{Grande:2006qi,Grande:2008re,Bauer:2010wj}.  Assume the existence of a DE component $X$ which exchanges energy with the running vacuum.   One possibility for $X$ would be a scalar
field $\chi$, but we are not going to make any specific assumption on the ultimate nature of $X$, e.g.  it could be an effective representation of dynamical fields of various sorts, or even the  behavior of higher order curvature terms in the effective action.  Since we still have the RVM, this means that we have a running vacuum term, so the total DE density,  $\rho_{\rm DE}$,  is given by the sum  $\rv+\rX$,  and the overall DE conservation law reads
\begin{equation}\label{conslawDE2}
\dot{\rho}_{\rm vac}+\dot{\rho}_X+\,\aX\,\rX\,H=0\,, \ \ \ \ \
\aX\equiv 3(1+\wX)\,,
\end{equation}
where $\wX$ is the EoS of $X$, which for simplicity we will assume constant.   In the previous conservation law it was assumed that $\Pv=-\rv$. Despite we now know that the running vacuum may depart slightly from that classical relation, we will still take it to hold as in \cite{Grande:2006nn}  since this feature does  not affect the main purposes of that study.  The departure of the EoS from $-1$ will, however, be important for the considerations of the $w$XCDM model discussed in the next section, which are of different nature despite it sharing the $X$ component in both cases.
For running $\rv$,  Eq.\,(\ref{conslawDE2})  shows that the
dynamics of the vacuum and the new component $X$ become
entangled. One can derive  the above conservation law starting from
the total energy-momentum tensor of the mixed DE fluid, with
$4$-velocity $U_{\mu}$:
\begin{equation}\label{TDE}
T^D_{\mu\nu}=T^{\rm vac}_{\mu\nu}+T^X_{\mu\nu}=
(\rv-\wX\,\rX)\,g_{\mu\nu}+(1+\wX)\rX\,U_{\mu}\,U_{\nu}\,.
\end{equation}
Then one can use the FLRW metric
\begin{equation}\label{FLRWm}
  ds^2=dt^2-a^2(t)\left(\frac{dr^2}{1-K\,r^2}
+r^2\,d\theta^2+r^2\,\sin^2\theta\,d\phi^2\right)
\end{equation}
and straightforwardly compute
$\bigtriangledown^{\mu}\,{T^D}_{\mu\nu}=0$, which yields \eqref{conslawDE2}.  The corresponding Friedmann's equation reads
\begin{equation}\label{eq:FriedmannLXCDM}
H^{2}\equiv \left( \frac{\dot{a}}{a}\right) ^{2}=\frac{8\pi\,G }{3}%
\left( \rmr +\rD\right) -\frac{K}{a^{2}}=\frac{8\pi\,G }{3}%
\left( \rmr +\rv+\rX\right) -\frac{K}{a^{2}}\,,
\end{equation}
and  for  the time
derivative of $H$ one finds
\begin{equation}\label{dH}
\dot{H}=-\frac{4\pi\,G}{3}\,[\amr\,\rmr+\aX\rX]+\frac{K}{a^2}\,.
\end{equation}
In practice we can simply set $K=0$ (flat three-dimensional spacetime).

The conservation law \eqref{conslawDE2} concerns the covariant conservation of the total composite DE density, namely the sum of the  running VED -- expressed by the RVM formula \eqref{VacuumEnergyDensityScalarField}-- and the new component $X$.  There is no exchange between DE and matter in this setup, hence  matter is conserved in the usual way, $\rM\sim a^{-3}$.  In these conditions the background cosmological equations of the $\CC$XCDM model can be fully solved analytically, see \cite{Grande:2006nn}\footnote{See also \cite{Grande:2006qi} for a variant $\CC$XCDM scenario, in which both matter and $X$ are conserved whereas the VED is running at the expense of another running quantity: the gravitational coupling $G$ Here, however, we will focus exclusively on the original scenario of  \cite{Grande:2006nn} at fixed $G$}.  We shall not discuss these cumbersome equations here, but we would like at least to stand out the main reason for studying the
$\CC$XCDM model in that reference, and hence the reason for the presence of the extra component $X$, which is to provide a possible explanation for the ``cosmic coincidence problem''\,\cite{Steinhardt1997}.
Recall that this problem is related to the behavior of the ratio
$r(z)\equiv\rD(z)/\rM(z)$  (``cosmic coincidence ratio'') between the dark energy density and the
matter energy density.  In the standard $\CC$CDM model, that ratio
unstoppably with the cosmic evolution because $\rD$ is
just the strictly constant vacuum energy density $\rvo$, whereas
$\rM\to 0$ with the expansion, so there is no apparent reason why we
should find ourselves precisely in an epoch where $r={\cal O}(1)$. This is the purported ``cosmic coincidence'' alluded to by the mentioned coincidence problem.
If it is a problem at all\,\cite{Ishak2005}, it has no known solution within the standard  $\CC$XCDM model.
However, it can be highly alleviated in the context of the
$\CC$XCDM model. This is so  thanks to the presence of the new DE component  $X$, which can prevent the aforementioned  cosmic coincidence ratio to increase forever. In fact,  one can easily understand from Eq.\eqref{eq:FriedmannLXCDM} that  for $\rX<0$ the function $r(z)$ will present a maximum at some redshift since we cannot have $H<0$. From explicit calculation using the analytical solution of the $\CC$XCDM model one finds
\,\cite{Grande:2006nn}:
\begin{equation}\label{zs}
z_{\rm max}=\left[\frac{\OL^0-\nu}{\wX\,(\Omega_X^0+\nu\,\OM^0)-
\epsilon\,(1-\OL^0)}\right]^{\frac{1}{3\,(1+\wX-\epsilon)}}-1\,,
\end{equation}
where we have defined $\epsilon\equiv\nu(1+\wX)$. One can show that
this point cannot be in our past, so it must generally lie
in the future (hence $-1<z_{\rm max}<0$). Beyond the maximum, there is a turning point $z_s<z_{\rm max}$ where there is a bounce, namely the
universe stops and reverses its evolution.  At the stopping point,
$\Omega_D(z_s)=-\OM(z_s)<0$. This is possible because we can have
$\Omega_X<0$ in the $\CC$XCDM.  Let us also notice that one can also achieve this situation with $\Omega_X>0$ and negative
cosmological constant  $\OL<0$.  The cosmic sum rule of the $\CC$XCDM allows this occurrence since today's  values of the cosmological parameters must satisfy
\begin{equation}\label{sumrule0}
\OMo+\ODo+\OKo=\OMo+\OLo+\OXo+\OKo=1\,.
\end{equation}
Mind, however, that  in the $\CC$XCDM model the  cosmological parameter associated to the physical cosmological term is not $\OLo$ but $\ODo=\OLo+\OXo$ since in this context the total VED is associated with the measurement of $\OD$, not with $\OL$.  In fact, it is not possible to dissociate $\OL$ from $\OX$ without further information, as only their sum $\OD$ can be determined at each point of the cosmic expansion.

\begin{figure}[t]
  \begin{center}
      \resizebox{0.7\textwidth}{!}{\includegraphics{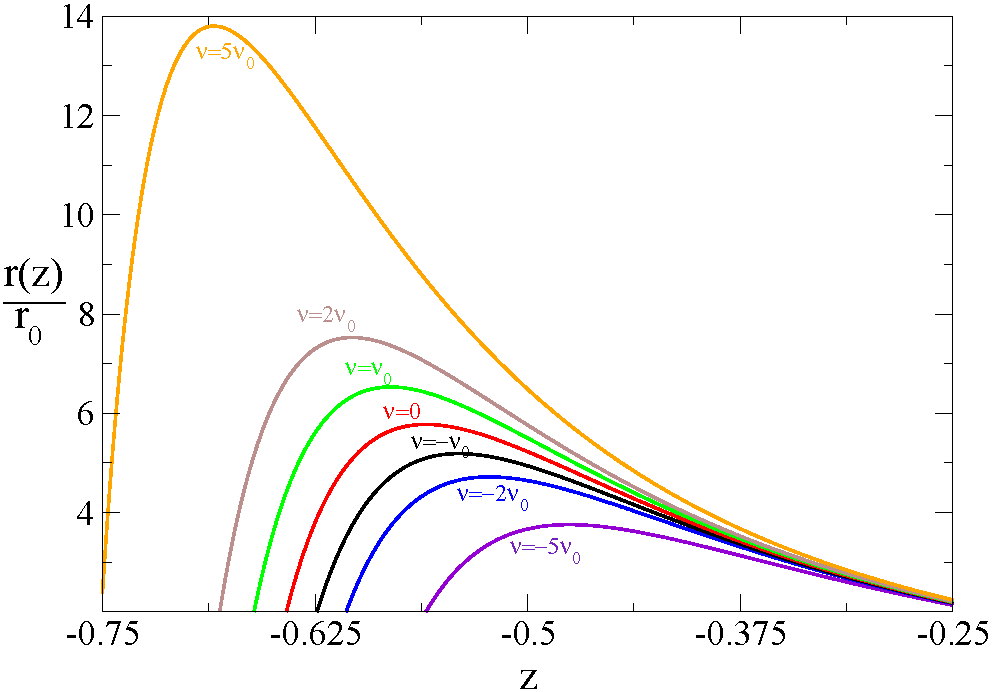}}
      \end{center}
    \caption{The cosmic coincidence ratio $r=\rD/\rM$, normalized to its current value $r_0$, for
$\wX=-1.85$, $\OMo=0.3,\OLo=0.75, \OXo=-0.05$ (flat space) and
different values of $\nueff$. Notice that  $\OXo<0$ insures stopping and
bouncing at a future point ($z=z_s<0$). All the maxima at the returning points satisfy $\,r_{\rm
max}<{\cal O}(10)$, which suggests a viable solution to the cosmic
coincidence problem. The values of $\nu$ are taken to be multiples
of $\nu_0\equiv 1/(12\pi)$.  Notice that even for $\nu=0$ we can alleviate the cosmic coincidence problem for $\OXo<0$\,\cite{{Grande:2006nn}}.}
\label{fig:ratio}
\end{figure}

The following observation is now in order: notice, from
eq.\,(\ref{conslawDE2}), that for $\nu=0$ (which implies no running of
the VED, i.e. $\dot{\rho}_{\rm vac}=0$ ) the  energy density of the new component $X$ is conserved; so, if in addition
$\Omega_X^0=0$ (vanishing current density of $X$), we must then have
$\Omega_X(z)=0$ at all times. Thus, in the limit
$\nu=\Omega_X^0=0$ we exactly recover the standard $\CC$CDM model.
And it is precisely in this limit that the formula (\ref{zs}) for
the redshift location of the maximum of the ratio $r(z)$ ceases to
make sense, as it is evident by simple inspection of that formula.
In contrast, when we make allowance for the parameter space of the
$\CC$XCDM model, the maximum for $r(z)$ does indeed exist and moreover the ratio
$r(z)$ stays bounded -- typically $r<{\cal O}(10)$ -- for the entire
cosmic history. This is manifest in the numerical examples exhibited  in Fig.\,\ref{fig:ratio}, for
various values of the parameters.  Therefore, the $\CC$XCDM model
can provide a nice solution to the cosmic coincidence problem
without deviating exceedingly from the $\CC$CDM model.    Furthermore, the detailed analysis of the cosmic perturbations in the $\CC$XCDM model indicate excellent
compatibility with the present observational data on structure
formation\,\cite{Grande:2008re}.

\section{\wXCDM: a daring cocktail of quintessence and phantom matter}\label{sec:wXCDM}

In the Introduction we have mentioned some studies devoted to the cosmological tensions from various perspectives, which include e.g.  the use of late DE transitions of the EoS, changes of the absolute magnitude of the SnIa and other calibrators at different rungs of the cosmic scale, smooth deformation models of the H(z) function etc.  For the most acute tension, the $H_0$ one, it  is generally believed that it is  primarily due to the discrepancy between local measurements and early-time observations\,\cite{Abdalla:2022yfr}.  More recent studies suggest, instead,  that  the core of the tension may originate between distance ladder measurements and
all other methods\,\cite{Perivolaropoulos:2024yxv}. Alternatively, it is proposed that it is between the Planck CMB measurements  and the first two-rung local distance ladders\cite{Huang:2024gfw}{.  It is probably too premature to reach any firm conclusion at present. In the meantime, it has  been argued in the literature that early-time new physics alone is insufficient to solve the $H_0$ tension\,\cite{Vagnozzi:2023nrq}, see also \cite{Gomez-Valent:2023uof}.


\begin{figure}[!t]
\begin{center}
\includegraphics[scale=0.07]{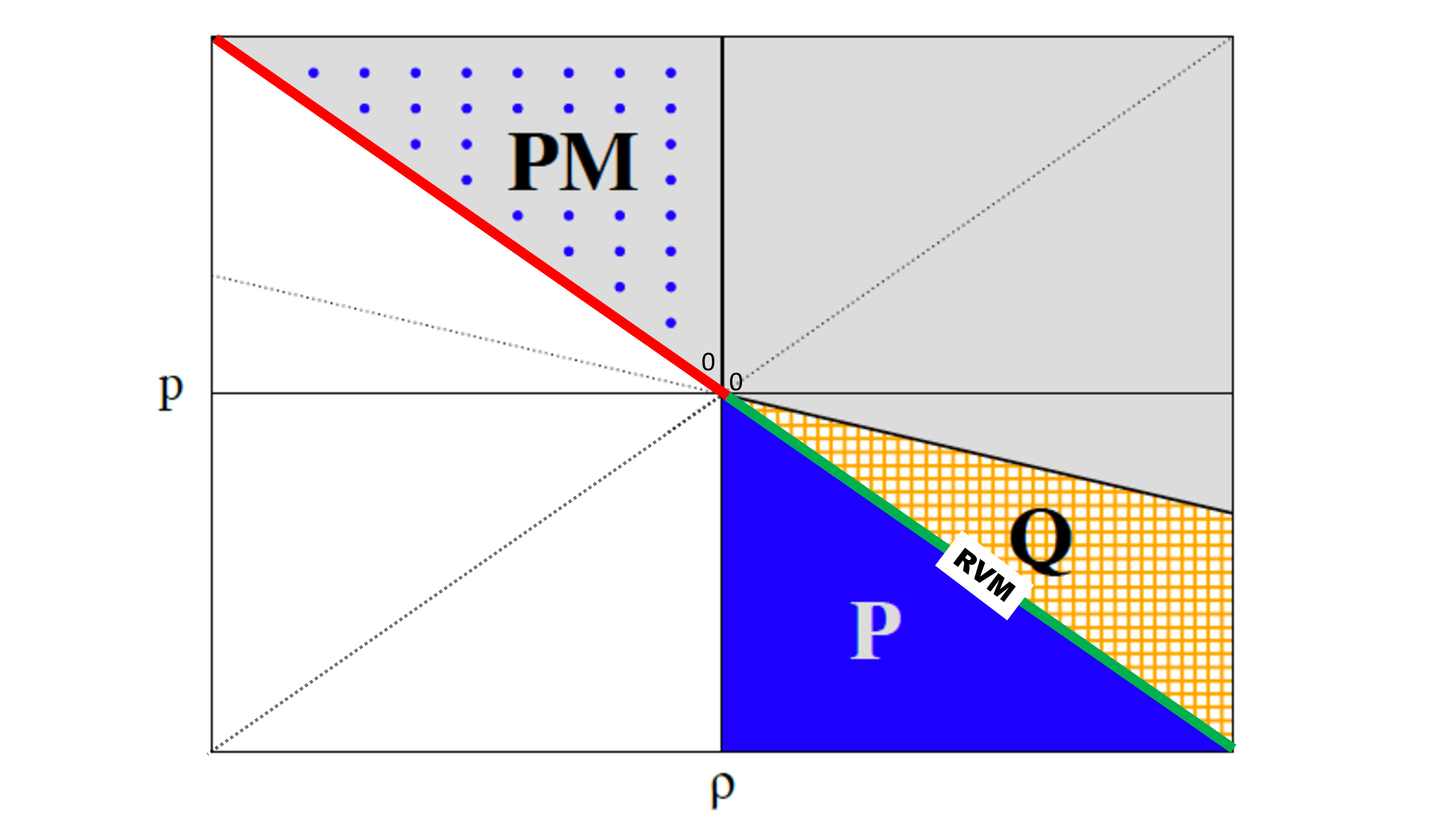}
\end{center}
\caption{\scriptsize EoS diagram describing a number of energy conditions for the cosmic fluids.   In the figure, (Q) denotes the  quintessence region: $-1<w<-1/3$ with $\rho>0$, $p<0$ (cross-hatched region);  (P) marks off the conventional phantom region: $w\le-1$ with $\rho>0$, $p<0$ (in blue). The gray-dotted sector corresponds to the  ``Phantom
Matter'' (PM):  $w<-1$ with $\rho<0$ and  $p>0$.  PM satisfies the strong energy condition: $\rho+p\geq 0, \ \rho+3p\geq 0$ (all of  the gray area) {but not the weak energy condition}: $\rho\geq 0, \ \rho+p\geq 0$ (shaded area, except regions P and PM). The DE component $X$ in the $w$XCDM model behaves as PM, whereas  $Y$ behaves effectively as quintessence. The EoS line $w=-1$ (phantom divide)  stands for  the classical vacuum. The RVM departs from this line owing to quantum effects\,\cite{CristianJoan2022b}.}
\label{fig:EC}
\end{figure}

 The foregoing condition is indeed satisfied by the model under consideration, i.e. the  $w$XCDM model\,\cite{wXCDM}, for which the physics of the early universe is left unchanged and moreover a composite structure of the DE  is favored in more recent epochs.  Here I will summarize the basics of this model, which can provide an exceptional fit to the cosmological data. We refer the reader to the quoted reference for more details.  The $w$XCDM model is not just an arbitrary deformation of the expansion history since it stems from the general idea of the running vacuum dynamics, which is grounded on QFT and string theory\,\cite{JSPRev2022,ReviewNickJoan2021}. It is in part  inspired from the structure of the $\CC$XCDM described in the previous section, although the main aim now  is no longer the cosmic coincidence problem but to provide a solution to the cosmic tensions explained in the Introduction, both the $H_0$ tension and the growth tension, and even to possibly account for the recent  JWST data on unexpected supermassive galaxies at high redshifts.  The two models share the $X$ component, but  in the \wXCDM the running $\CC$ is superseded  by another dynamical component $Y$  whose EoS,  $w_Y$, can be different from $-1$. This is justified since, as we have indicated in the previous section, the running vacuum has indeed an effective EoS which departs slightly from  $-1$ owing to quantum effects\,\cite{CristianJoan2022b}.  The \wXCDM acts as  a simplified implementation  of the full features encoded in the  $\CC$XCDM but the former can effectively mimic the main features of the latter.  Furthermore, while in the  $\CC$XCDM the two components (the running $\CC$ and the new component $X$) are present at the same time across the cosmic evolution, in the \wXCDM they act in sequence. To be precise, the component  $X$ enters only above a transition redshift $z>z_t$ (fitted from the data) and $Y$ enters below that redshift until the current time.  In actual fact, the action of the phantom matter (PM) component $X$ is only limited to a  ``bubble of spacetime'' owing to a dynamical mechanism which kills the false vacuum (`` phantom vacuum'')\,\cite{PhantomVacuum2021}.   In the subsequent section we summarize the nature of the PM component by considering the different types of energy conditions for the cosmic fluids.

\subsection{Phantom matter in the context of the energy conditions}\label{sec:phantomvacPM}

For the sake of a better contextualization of the PM option within the class of energy conditions, we display  in  Fig. \ref{fig:EC} some of the most common possibilities for the EoS satisfied by the cosmic fluids. In it, we can see e.g. the locus of the  phantom matter (PM) fluid. As can be seen, it is very different from the usual phantom DE, it lies actually in its antipodes.  This must be stressed, these two phantom-like components are dramatically disparate. The existence of the aforementioned two different  phases of the DE ($X$ and $Y$)  in our universe separated by a transition redshift $z_t$, and in particular the possibility that $X$ embodies a PM phase,  is not just a whimsical assumption since it is found in theoretical contexts such as the stringy RVM approach\,\cite{PhantomVacuum2021}, which points to the existence of transitory domains or bubbles of PM when the universe asymptotes to a de Sitter epoch.  The phantom-like component $X$  actually behaves as PM in the \wXCDM.   This entails negative energy density ($\Omega_X=\rX/\rho_c<0$) and hence positive pressure ($p_X>0$).   The behavior of PM is, as indicated,  far  away from that of the usual phantom DE.  For, despite they both satisfy $w\lesssim-1$,  PM produces positive pressure at the expense of negative energy density, and hence it resembles usual matter (therein its name!\cite{Grande:2006nn}). This is in stark contrast to phantom DE, which produces negative pressure with positive energy. `Conventional' phantom DE  has been amply considered in the literature to describe different features of the DE, including a possible explanation for the cosmological tensions on  $H_0$ and structure formation in a variety of  frameworks (see the Introduction). However, in  our approach,  conventional phantom DE is not used at all.  As a matter of fact, phantom DE is not favored in our analysis of the cosmological data, only PM is singled out  as the best option  for a possible alleviation, or  even  full resolution,  of the cosmological tensions.

While the PM component  $X$ acts first during the cosmic expansion, restricted to a bubble of spacetime,  the  $Y$ component takes its turn in the most recent universe (below $z_t\sim 1.5$) up to our days, see Table 1.  The $Y$ component is more conventional, we have $\Omega_Y>0$ and  $p_Y<0$, and hence it could perfectly behave as quintessence or phantom DE. Nevertheless,  it is the numerical fitting to the cosmological data which must decide phenomenologically which one of these two options is more favored.  The result is that its EoS satisfies $w_Y\gtrsim-1$  (cf. Table 1 and  Fig.\,\ref{fig:wXwY}), hence $Y$  is quintessence-like.  The transition redshift $z_t$ separating the two components $X$ and $Y$ is also determined by the fitting procedure  (cf. Table 1).  Therefore, the characteristic free parameters of model $w$XCDM are just three, the transition point and the two EoS parameters of the DE components:  $(z_{t}, w_X, w_Y)$.   Let us clarify that the  density parameters for the DE components are not free since e.g. the  value of $\Omega_Y^0\equiv\Omega_Y(z=0)$ depends on the fitting values of $H_0, \omega_b, \omega_{\rm dm}$ (cf. Table 1).
In addition, the respective values of $\Omega_X$ and of $\Omega_Y$  immediately above and below $z_{t}$ are assumed to be equal in absolute value:  $|\Omega_X(z)|=\Omega_Y(z)$ at $z=z_{t}$. This kind of assumption aims at reducing the number of parameters.  We cannot exclude that by relaxing this assumption it may further improve the fit quality, but it would imply more modelling features which we wish to avoid at this point.

In \cite{PhantomVacuum2021} it is argued that the PM-dominated era can be present in the early cosmic evolution near a de Sitter phase, but new episodes  could also reflourish in the late universe. As shown in\cite{BasMavSol}, the presence of chiral matter fields at the exit from inflation enables the cancellation of the gravitational anomalies and  at large scales the universe can recover its normal FLRW background profile.  Because the chiral matter fields will get more and more  diluted with the cosmic expansion,  this will entail an incomplete cancellation of  the gravitational  anomalies, and as a result these may eventually re-surface  near  the  current quasi-de Sitter era.  For this reason it is conceivable to admit the  formation of phantom matter bubbles or domains tunneling   into our universe at relatively close redshift ranges before the ultimate de Sitter phase takes over in the late universe.  These bubbles of PM are characterized by  positive pressure $p>0$ (see Fig.\,\ref{fig:EC}) and therefore could be the seed of  structure formation at unusually high redshifts. If so, this could explain on fundamental grounds the overproduction of large scale structures at unexpected places and times deep in our past.   This ideology provides an excellent framework for a fit to the overall cosmological data which proves to be (far) better than that of the standard $\CC$CDM model (cf. Table 1).

\subsection{Fitting the modern cosmological data with the \wXCDM}\label{sec:fittingwXCDM}

To constrain the \wXCDM parameters, we make use of the following cosmological data sets (see \cite{wXCDM} for a more detailed description and detailed references for all of them):

\begin{itemize}
\item The full Planck 2018 CMB temperature, polarization and lensing likelihoods.

\item The SnIa contained in the Pantheon+ compilation,  calibrated with the cosmic distance ladder measurements of the SH0ES Team.

\item  $H(z)$ data  from cosmic chronometers (CCH).

\item Transverse (aka angular or 2D) BAO data from Refs. \cite{BAO2D}.  These data deserve a particular consideration.  In fact, transverse BAO  data are claimed to be less subject to model-dependencies, since they are obtained without assuming any fiducial cosmology to convert angles and redshifts into distances to build the tracer map. These data points are extracted from the two-point angular correlation function or its Fourier transform. In anisotropic (or 3D) BAO analyses, instead, a fiducial $\CC$CDM  cosmology is employed to construct the 3D map in redshift space, hence potentially introducing model-dependencies. Furthermore, some studies have suggested that uncertainties may be underestimated by roughly a factor of two in these analyses \cite{Anselmi:2018vjz}. In contrast to the 3D BAO data, 2D BAO observations still leave room  for low-redshift solutions to the Hubble tension while respecting the constancy of the absolute magnitude of SnIa, which is the kind of approach that we propose here.

\item The data on large-scale structure at $z\lesssim 1.5$.  For these data we also wish to add some qualification. These are LSS observations on the weighted growth
 $f(z)\sigma_8(z)$, with $f(z)=-(1+z)d\ln\delta_m/dz$ the linear growth rate, $\delta_m=\delta\rho_m/\rho_m$ is  the matter density contrast, and $\sigma_8(z)$ the rms mass fluctuations at a scale $R_8=8h^{-1}$ Mpc, see Table 3 of  \cite{SolaPeracaula:2023swx}. These data points, however, are taken  using  fiducial cosmology with $h\sim 0.67$, which translates into measurements at a characteristic length scale of $R\sim 12$ Mpc. At variance with this rather common practice,  and as noted already in the Introduction, we adhere instead to the alternative procedure advocated in\cite{Sanchez:2020vvb,eBOSS:2021poy,Semenaite:2022unt}, and therefore we treat these LSS observations as data points on $f(z)\sigma_{12}(z)$, using a Fourier-transformed top-hat window function $W(kR_{12})$  in the computation of $\sigma_{12}(z)$. {The advantage   is that the scale $R_{12}=12$ Mpc is independent of the parameter $h$.  Following this approach we find more consistent results for the growth of matter fluctuations. }
\end{itemize}

Using the above data sources for our fit, we present our main numerical results  in Table 1 and in the triangle plots of Figs. \ref{fig:triangle_plot_in}-\ref{fig:wXwY}.

\begin{table}
\resizebox{\textwidth}{!}{
\begin{tabular}{|c ||c | c  |c |}
\hline
{\small Parameter} & {\small $\Lambda$CDM}   & {\small $w$XCDM }  & {\small $\Lambda_s$CDM}
\\\hline
$\omega_b$ & $0.02281\pm 0.00014$ (0.02278) & $0.02241\pm 0.00013$ (0.02260) & $0.02236^{+0.00016}_{-0.00018}$ (0.02232) \\\hline
$\omega_{\rm dm}$ &  $0.1153\pm 0.0009$ (0.1148)   & $0.1199\pm 0.0010$ (0.1196) & $0.1205^{+0.0015}_{-0.0016}$ (0.1216) \\\hline
$\ln(10^{10}A_s)$ & $3.066^{+0.016}_{-0.018}$ (3.080) &  $3.037\pm 0.014$ (3.034)  & $3.036^{+0.015}_{-0.016}$ (3.030) \\\hline
$\tau$ & $0.069^{+0.008}_{-0.010}$ (0.076)  & $0.051\pm 0.008$ (0.048) &  $0.050^{+0.008}_{-0.009}$ (0.046)   \\\hline
$n_{s}$ &  $0.978\pm 0.004$ (0.981) &   $0.967\pm 0.004$ (0.969) &  $0.966^{+0.004}_{-0.005}$ (0.961)  \\\hline
$H_{0}$ [km/s/Mpc] &  $69.82^{+0.41}_{-0.44}$ (70.05) &  $72.75^{+0.57}_{-0.71}$ (72.36) & $72.24^{+0.99}_{-0.75}$ (73.82)\\\hline
$z_t$ & $-$ & $1.46^{+0.02}_{-0.01}$ (1.47) & $1.61^{+0.22}_{-0.18}$ (1.47) \\\hline
$w_X$ & $-$  & $-1.16^{+0.13}_{-0.16}$ (-1.16)  &  $-$ \\\hline
$w_Y$ & $-$  & $-0.90\pm 0.03$ (-0.88)&  $-$ \\\hline\hline
$\Omega_m^0$ & $0.283\pm 0.005$ (0.280) & $0.269\pm 0.005$ (0.272)  &  $0.267\pm 0.005$ (0.264)\\\hline
$M$ &  $-19.372^{+0.011}_{-0.012}$ (-19.362)   & $-19.273^{+0.015}_{-0.016}$ (-19.282) & $-19.278^{+0.026}_{-0.020}$ (-19.261)   \\\hline
$\sigma_{12}$ & $0.780\pm 0.007$ (0.884) &  $0.776\pm 0.007$ (0.772) & $0.782^{+0.007}_{-0.006}$ (0.784)
 \\\hline\hline
$\chi^2_{\rm min}$ &  $4166.76$  & $4107.62$ &  $4120.04$ \\\hline
$\Delta$DIC &  $-$  & $57.94$  & $40.16$  \\\hline
$\Delta$AIC &  $-$  & $53.14$  & $44.72$\\\hline
\end{tabular}}
\vspace{0.3cm}

{\scriptsize Table 1. Fitting results. Mean values and uncertainties at 68\% CL obtained with the full data set CMB+CCH+SnIa+SH0ES+BAO+$f\sigma_{12}$ used in \cite{wXCDM}.  Best-fit values are shown  in brackets.  In the last three lines, we display the values of the minimum $\chi^2$ and the differences in the standard information criteria DIC and AIC.  Positive values of $\Delta$DIC and $\Delta$AIC indicate preference for the new models over the $\CC$CDM.  The preference is  high  both for the \wXCDM\cite{wXCDM} and $\CCS$\cite{Akarsu:2023mfb} models, but the former is much more preferred since it differs from the latter by about 10 points of AIC and more than 17 points of  DIC. The extraordinary quality of the \wXCDM fit is unprecedented for the data used. }
\end{table}

\begin{figure}[t!]
    \begin{center}
    \includegraphics[scale=0.25]{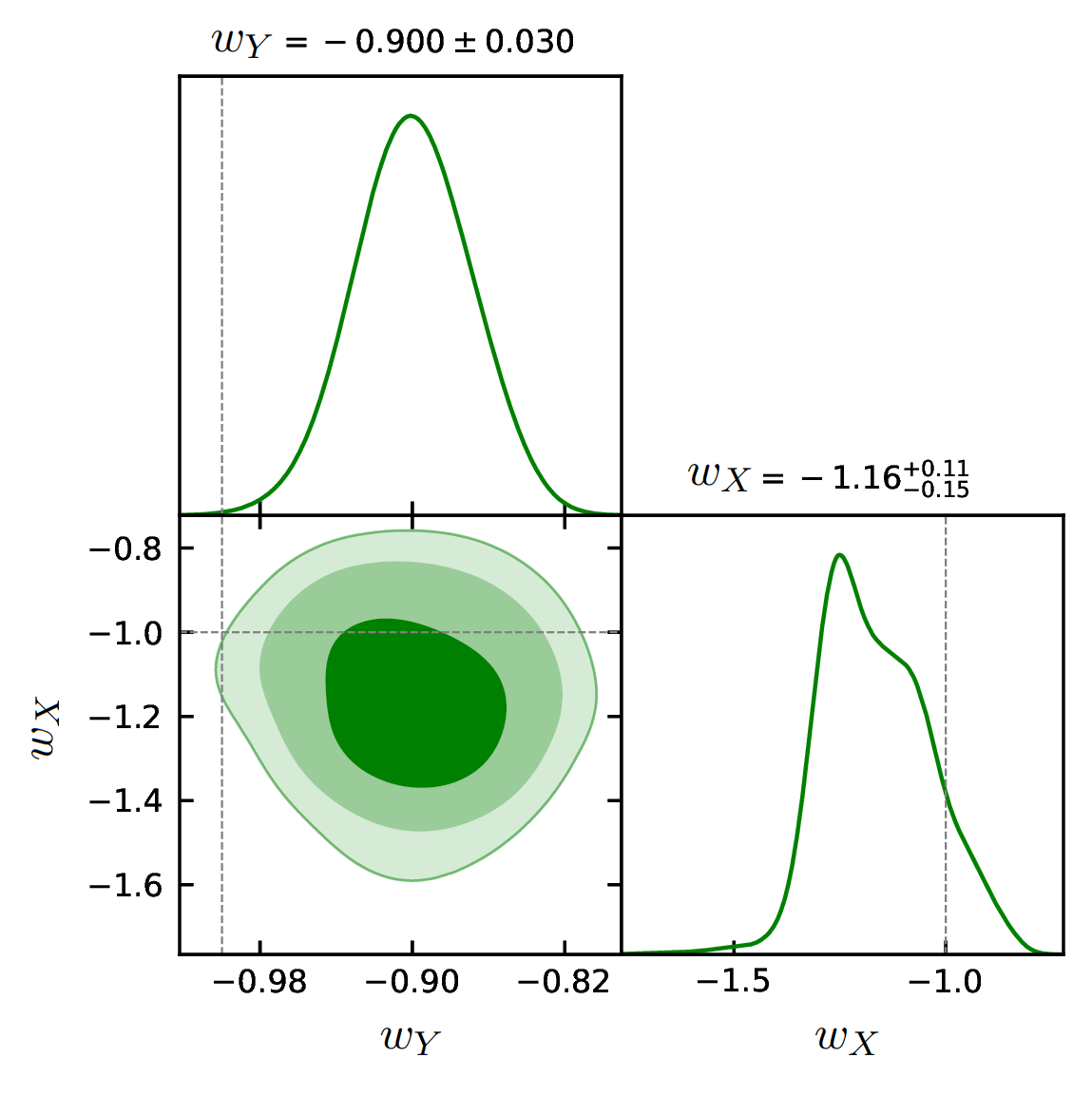}
    \end{center}
    \caption{\scriptsize Confidence regions in the EoS plane  of the $w$XCDM model, and the corresponding one-dimensional posterior distributions. The dotted lines correspond to $w_X=-1$ and $w_Y=-1$. The intersection of the horizontal and vertical line pinpoints the $\Lambda_s$CDM model\cite{Akarsu:2023mfb}, which can be seen to fall $\gtrsim 3\sigma$ away from the preferred region of the $w$XCDM.  There is a clear preference for $w_X<-1$ (phantom matter, since $\OX<0$) and $w_Y\gtrsim-1$ (quintessence). As a result, among the models considered only the \wXCDM is consistent with the recent DESI data\cite{DESI:2024mwx}.}
    \label{fig:wXwY}
\end{figure}

\begin{figure}[t!]
    \begin{center}
    \includegraphics[scale=0.12]{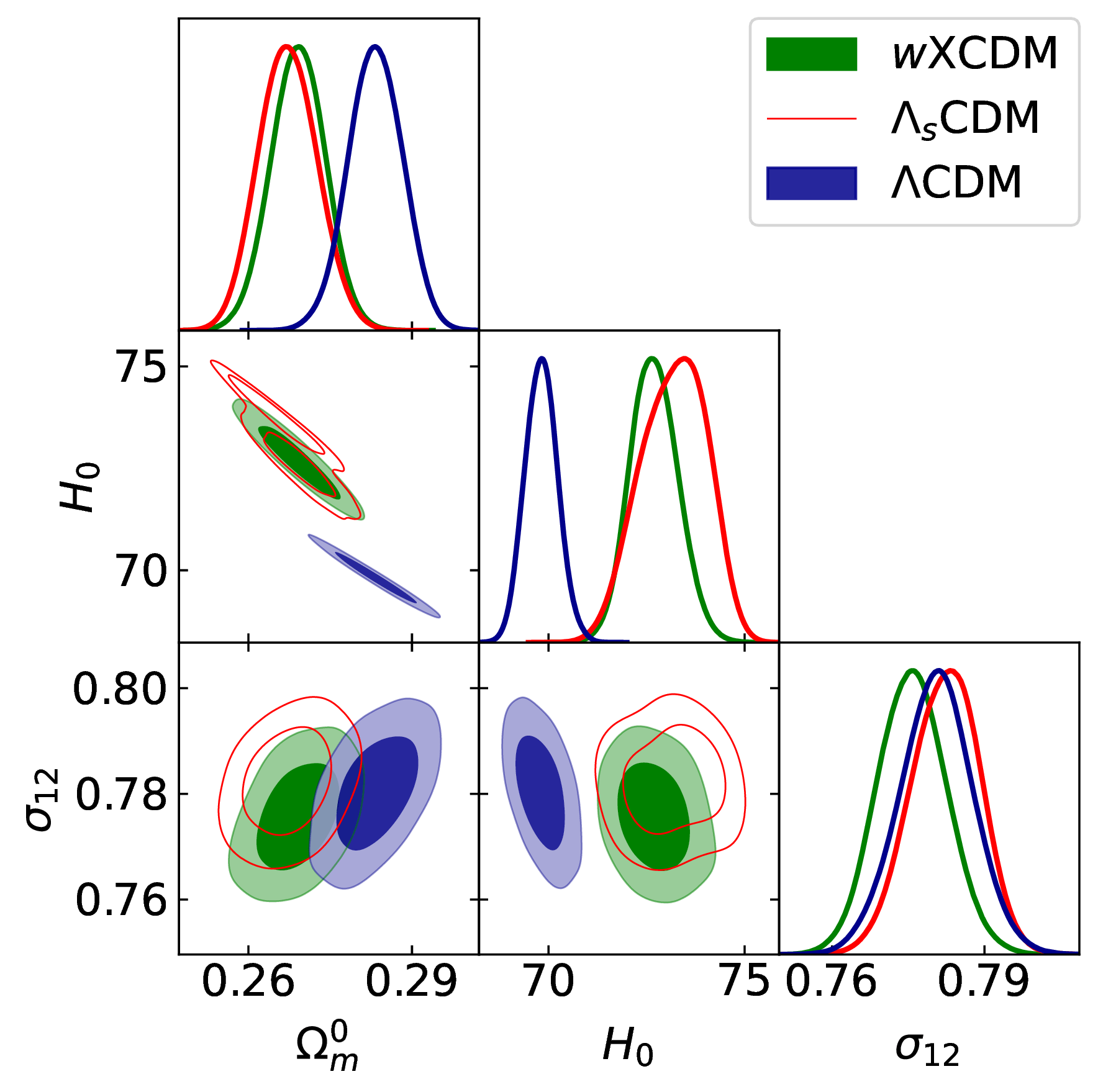}
    \end{center}
    \caption{\scriptsize Contour plots for the various models corresponding to $68\%$ and $95\%$ c.l.  together with  the associated one-dimensional posterior distributions for some of the parameters which are relevant in the current discussion of the cosmological tensions.  $H_0$ is given in km/s/Mpc. The complete triangle plot is presented in the appendix of \cite{wXCDM}.}
    \label{fig:triangle_plot_in}
\end{figure}

Let us briefly summarize why the \wXCDM can deal effectively with the cosmological tensions.  Let us consider $H_0$ first.  Recall that  $\theta_*$ (the angular size of the sound horizon)   is given by $ \theta_*=r_d/D_A(z_*)$, where $r_d$ is the sound horizon at decoupling and $D_A(z_*)$ is the angular diameter distance at this point: $D(z_*)=c(1+z)^{-1}\int_0^{z_*} \frac{dz'}{H(z')}$.  As indicated, we shall not modify the physics prior to decoupling (i.e. for  $z>z_*$) and therefore we aim essentially at changes occurring after recombination.  However, these changes cannot modify the extremely well known value  of $\theta_* $.  This riddle can be solved by implementing internal changes in $D_A(z_*)$ that compensate each other but that leave a measurable imprint on the late universe (e.g.  producing a change of the current value of the  Hubble parameter   $H_0$). In fact,  this can be accomplished as follows.  Notice that  the integral  defining $D_A$ can be split at the transition point $z_t$ which separates the two regimes (PM for $z>z_t$ and quintessence for $z<z_t$):
\begin{equation}\label{eq:integral}
 \int_0^{z_*} \frac{dz'}{H(z')}=  \int_0^{z_t} \frac{dz'}{H_{\rm Q}(z')}  + \int_{z_t}^{z_*} \frac{dz'}{H_{\rm PM}(z')}\,,
\end{equation}
where $H_{\rm Q}$ and  $H_{\rm PM}$  are the Hubble functions within the quintessence (Q) and phantom matter (PM) regimes depicted in Fig.\,\ref{fig:EC}.
Using this composite RVM model  we can easily explain  within our PM scenario (based on the  \wXCDM model)  why $H_0$ is found larger than in the $\CC$CDM.  It is important to realize that because the energy of the $X$ entity is negative (and non-negligible at high redshift, in absolute value), this enforces $H_{\rm PM}$ to be smaller than in the standard regime and hence the second integral in \eqref{eq:integral} is larger than the usual contribution in the $\CC$CDM. This enforces  a smaller value of the first integral on the r.h.s. of the above expression  and hence a larger value of the expansion rate $H$ for $z<z_t$, This  is possible in the quintessence stage characterized by $H_{\rm Q}$ provided we  have a larger value of $H_0$ (to be fixed by our fit) . In this way the two changes can compensate each other and  $D_A(z_*)$  (and $\theta_*$)  remains the same without modifying the physic of the early universe, i.e. the value of $r_d$.
 Quantitatively, the parameter $H_0$ emerging from our fit  (cf. Table 1) is in full agreement with that of  SH0ES ($H_0=73.04\pm 1.04$ km/s/Mpc)\cite{Riess:2021jrx} to within $\sim  0.25\sigma$. The Hubble tension is therefore completely washed out (see also  Fig. \ref{fig:triangle_plot_in}).  At the same time the rate of LSS formation becomes suppressed below $z_{t}\sim 1.5$ during the quintessence-like regime, which is in accordance with the observations. This can be easily understood by throwing a glance to the equation for the matter density contrast $\delta_m$ in the \wXCDM model. Before the transition at $z_{t}$ (i.e. at $z>z_{t}$), the matter perturbations equation reads
\begin{equation}\label{eq:dcX}
\delta^{\prime\prime}_m+\frac{3}{2a}\left(1-\Omega_X(a)w_X\right)\delta^\prime_m-\frac{3}{2a^2}(1-\Omega_X(a))\delta_m=0\,,
\end{equation}
where the primes denote derivatives with respect to the scale factor.  Since PM has negative energy density ($\Omega_X<0$) and positive pressure (due to $w_X<-1<0$)  it induces a decrease of the friction term and an increase of the Poisson term in Eq. \eqref{eq:dcX}. Both effects are cooperative and cause an enhancement of the structure formation processes in the PM bubbles. Notice also that this would not occur for ordinary phantom DE or quintessence (cf. Fig.\,\ref{fig:EC}), for which the friction term gets enhanced and the Poisson term suppressed, i.e. just the opposite behavior  of  PM.

Focusing now on  the $w$XCDM parameters, in Fig. \ref{fig:wXwY} we show the constraints obtained in the EoS plane  $w_Y$-$w_X$, which involves two of the characteristic new parameters of the $w$XCDM model (the third one being $z_t$, shared with $\CC_s$CDM). The central value of $w_X=-1.16$ falls in the phantom-like region (in fact, PM region since $\Omega_X<0$) but is compatible with $-1$ at $\sim 1\sigma$ CL.  There is, in contrast, a non-negligible ($\sim 3.3\sigma$) preference for a quintessence-like evolution of the  DE for the low redshift range nearer to our time ($z<z_{t}$): $w_Y=-0.90\pm 0.03$. This situation is beautifully  consistent with the recent DESI results suggesting dynamical DE\,\cite{DESI:2024mwx}.  Besides, we can see in Table 1 and  in Fig. \ref{fig:triangle_plot_in} that the amplitude of the power spectrum at linear scales that is preferred by the data is optimal for the \wXCDM,  $\sigma_{12}\sim 0.77$.  This can again be understood by looking at the equation for the density contrast, whose form in the range  $z<z_{t}$ is identical to Eq. \eqref{eq:dcX}, but with the replacements $\Omega_X\to \Omega_Y$ and $w_X\to w_Y$.


\section{Conclusions and outlook} \label{sec:conclusions}

 In this short review presentation I have described  the idea of the running vacuum model (RVM)\,\cite{JSPRev2022} and some of its composite DE extensions, such as the $\CC$XCDM\,\cite{Grande:2006nn}  and the  \wXCDM\cite{wXCDM}. The RVM has a long story, see \,\cite{JSPRev2013} and references therein,  but only recently its vacuum structure was rigorously derived  from  QFT calculations in curved spacetime \cite{CristianJoan2020,CristianJoan2022a,CristianJoan2022b,CristianJoanSamira2023}. The RVM provides a smooth renormalization  of the vacuum energy density (VED) and reasserts its preeminent  status  in cosmology since the latter becomes a firm (and fundamental) candidate for describing the DE in the universe without appealing to \textit{ad hoc} DE  fields. The RVM quantum vacuum is dynamical and appears to be free from fine tuning diseases, which have been attributed too often and too exclusively to the cosmological constant $\CC$.  This is a mistake.  The RVM restores the existence of a theoretically solid $\CC$ as the ultimate cause for the speeding up of the universe, although it is no longer a rigid constant but a  dynamical quantity. The DE is dynamical because herein  the VED,  wherefore   $\CC$, acquire both  (quantum!) dynamics with the cosmic expansion: $\CC=\CC(H$).

On a more practical vein, we have considered a possible solution to the cosmological tensions within the \wXCDM\cite{wXCDM}, a simplified version of the old existing $\CC$XCDM framework\,\cite{Grande:2006nn,Grande:2006qi,Grande:2008re},  which was inspired in the theoretical context of the RVM. Our present formulation benefits  from the  theoretical developments made in the context of the stringy version of the running vacuum model (StRVM)\cite{ReviewNickJoan2021,PhantomVacuum2021}.  The DE component $X$ that is involved in the $\CC$XCDM and the \wXCDM  can play the role of phantom matter (PM). In the StRVM, the PM phase  is only transitory and is predicted to appear prior to the  eventual transition to de Sitter phase in the future.  The most salient feature of PM is that it acts as a temporary stage with negative energy and positive pressure. It need not be a pure anti-de Sitter (AdS) phase (this would be only a very particular situation)  as  its EoS $w_X$ is not necessarily $-1$, but it must satisfy $w_X\lesssim -1$.  By fitting the  $w$XCDM model to the modern cosmological data (using exclusively transversal BAO as an important ingredient), we find that the PM phase appears  above a transition redshift $z_{t}\simeq 1.5$ (cf. Table 1).  It remains confined into a bubble of spacetime.   However, below $z_t$ the PM phase ceases to occur and the field $Y$ (viz. the would-be running VED in the full $\CC$XCDM\,\cite{Grande:2006nn})  takes its turn  with an effective EoS which is found  to be of quintessence type ($w_Y\gtrsim -1$) at $\sim 3.3\sigma$ c.l., this being nicely consistent with the first DESI data release\cite{DESI:2024mwx}.
In the $w$XCDM, the formation of a PM bubble is induced by quantum fluctuations associated to the nearing of the universe towards a de Sitter phase\cite{PhantomVacuum2021}. This phenomenon need not be isolated in time: the bubbles of PM  could well be operating more than once at earlier times in the late universe, what would trigger an anomalous outgrowth of structures at even higher redshifts, say in the range  $z\sim 5-10$. This might explain the appearance of the large scale structures recently spotted at unusually high redshift by the JWST mission\,\cite{Labbe:2022ahb}.  Such LSS anomalies, which have no explanation in the $\CC$CDM,  might also be described within the current proposal in terms of PM bubbles of spacetime\cite{wXCDM2}.

\section*{Acknowledgements}
{\scriptsize
Work  partially supported by grants PID2022-136224NB-C21 and  PID2019-105614GB-C21, from MCIN/AEI/10.13039/501100011033.  I am funded also  by  2021-SGR-249 (Generalitat de Catalunya) and
CEX2019-000918-M (ICCUB, Barcelona). Networking support by the COST Association Action CA21136 ``{\it Addressing observational tensions in cosmology
with systematics and fundamental physics (CosmoVerse)}'' is acknowledged. This presentation  is  based in part on works with my collaborators  A.  G\' omez-Valent,  N.E. Mavromatos,  J. de Cruz P\'erez, C. Moreno-Pulido and S.  Cheraghchi.}

\end{document}